\documentclass[12pt,epsf]{article}
\usepackage{graphicx}

\usepackage{amssymb}

\begin{document}

\def\a{\alpha}
\def\b{\beta}
\def\c{\varepsilon}
\def\d{\delta}
\def\e{\epsilon}
\def\f{\phi}
\def\g{\gamma}
\def\h{\theta}
\def\k{\kappa}
\def\l{\lambda}
\def\m{\mu}
\def\n{\nu}
\def\p{\psi}
\def\q{\partial}
\def\r{\rho}
\def\s{\sigma}
\def\t{\tau}
\def\u{\upsilon}
\def\v{\varphi}
\def\w{\omega}
\def\x{\xi}
\def\y{\eta}
\def\z{\zeta}
\def\D{\Delta}
\def\G{\Gamma}
\def\H{\Theta}
\def\L{\Lambda}
\def\F{\Phi}
\def\P{\Psi}
\def\S{\Sigma}
\def\BR{{\rm Br}}
\def\o{\over}
\def\beq{\begin{eqnarray}}
\def\eeq{\end{eqnarray}}
\newcommand{\nn}{\nonumber \\}
\newcommand{\gsim}{ \mathop{}_{\textstyle \sim}^{\textstyle >} }
\newcommand{\lsim}{ \mathop{}_{\textstyle \sim}^{\textstyle <} }
\newcommand{\vev}[1]{ \left\langle {#1} \right\rangle }
\newcommand{\bra}[1]{ \langle {#1} | }
\newcommand{\ket}[1]{ | {#1} \rangle }
\newcommand{\EV}{ {\rm eV} }
\newcommand{\KEV}{ {\rm keV} }
\newcommand{\MEV}{ {\rm MeV} }
\newcommand{\GEV}{ {\rm GeV} }
\newcommand{\TEV}{ {\rm TeV} }
\def\diag{\mathop{\rm diag}\nolimits}
\def\Spin{\mathop{\rm Spin}}
\def\SO{\mathop{\rm SO}}
\def\O{\mathop{\rm O}}
\def\SU{\mathop{\rm SU}}
\def\U{\mathop{\rm U}}
\def\Sp{\mathop{\rm Sp}}
\def\SL{\mathop{\rm SL}}
\def\tr{\mathop{\rm tr}}

\newcommand{\bear}{\begin{array}}  
\newcommand {\eear}{\end{array}}
\newcommand{\la}{\left\langle}  
\newcommand{\ra}{\right\rangle}
\newcommand{\non}{\nonumber}  
\newcommand{\ds}{\displaystyle}
\newcommand{\red}{\textcolor{red}}
\def\ubl{U(1)$_{\rm B-L}$}
\def\REF#1{(\ref{#1})}
\def\lrf#1#2{ \left(\frac{#1}{#2}\right)}
\def\lrfp#1#2#3{ \left(\frac{#1}{#2} \right)^{#3}}
\def\OG#1{ {\cal O}(#1){\rm\,GeV}}



\baselineskip 0.7cm

\begin{titlepage}

\begin{flushright}
UT-11-27\\ 
IPMU 11-0133
\end{flushright}

\vskip 1.35cm
\begin{center}
{\large \bf
Higgs Mass and Muon Anomalous Magnetic Moment \\
in Supersymmetric Models with Vector-Like Matters
}
\vskip 1.2cm
Motoi Endo$^{(a)(b)}$, Koichi Hamaguchi$^{(a)(b)}$, Sho Iwamoto$^{(a)}$, Norimi Yokozaki$^{(a)}$
\vskip 0.4cm

{\it $^{(a)}$ Department of Physics, University of Tokyo,
   Tokyo 113-0033, Japan\\
$^{(b)}$ Institute for the Physics and Mathematics of the Universe (IPMU), \\
University of Tokyo, Chiba, 277-8583, Japan
}

\vskip 1.5cm

\abstract{
We study the muon anomalous magnetic moment (muon $g-2$) and the Higgs boson mass
in a simple extension of the minimal supersymmetric (SUSY) Standard Model with extra vector-like matters, in the frameworks of gauge mediated SUSY  breaking (GMSB) models
and gravity mediation (mSUGRA) models.
It is shown that the deviation of the muon $g-2$ and 
a relatively heavy Higgs boson can be simultaneously explained in large $\tan\beta$ region.
(i) 
In GMSB models, the Higgs mass can be more than 135 GeV (130 GeV) in the region where the muon $g-2$ is consistent with the experimental value at the $2\sigma$ ($1\sigma$) level,
while maintaining the perturbative coupling unification. 
(ii)
In the case of mSUGRA models with universal soft masses, the Higgs mass can be as large as about 130 GeV 
when the muon $g-2$ is consistent with the experimental value at the $2\sigma$ level.
In both cases, the Higgs mass can be above 140 GeV if the $g-2$ constraint is not imposed.
}
\end{center}
\end{titlepage}

\setcounter{page}{2}

\section{Introduction}
Supersymmetry (SUSY)~\cite{Martin:1997ns} is one of the best candidates for new physics beyond the Standard Model of particle physics. It protects the huge hierarchy between the electroweak scale and unification scales against the radiative corrections, and the particle content of the minimal SUSY Standard Model (MSSM) leads to a beautiful unification of the three gauge couplings of the Standard Model. The MSSM also gives a natural framework to break the electroweak symmetry radiatively. In addition, it contains a natural dark matter candidate as the lightest SUSY particle.

In this work, we address two of important phenomenological issues of the SUSY Standard Model, the muon anomalous magnetic moment (muon $g-2$) and the mass of the Higgs boson. Latest studies have reported that the discrepancy of the measured 
muon $g-2$~\cite{Bennett:2006fi} from the Standard Model prediction is more than $3\sigma$~\cite{g-2_hagiwara, g-2_davier}.
It is quite interesting that the low-energy SUSY Standard Model can naturally explain this discrepancy~\cite{Moroi:1995yh}. In this paper, the muon $g-2$ anomaly is considered as a signal of the low-energy SUSY.

On the other hand, one of the most remarkable predictions of the SUSY Standard Model is an upper bound on the lightest Higgs boson mass. This is of particularly vital importance in light of recent impressive progress in Higgs boson search at the LHC~\cite{HiggsEPS}. 
It is well-known that, in the low-energy MSSM, including the radiative corrections~\cite{mh1,mh2},
the mass of the lightest Higgs boson can be as large as, but not more than, about 130 GeV. 
It can be raised more by taking the soft masses of SUSY particles to be far above the electroweak scale (and giving up solving the hierarchy problem),  but then the muon $g-2$ is no longer explained by the SUSY.
Another possibility to increase the Higgs mass is the next-to-MSSM~\cite{Ellwanger:2009dp}. However, as far as the theory is assumed to remain perturbative up to the unification scale, a sizable increase of the Higgs boson mass can be obtained only in small $\tan\beta$ region, where the SUSY contribution to the muon $g-2$ is small.
A natural question is, therefore, how large the Higgs boson mass can be while keeping the SUSY explanation of the muon $g-2$ anomaly and the perturbative coupling unification.

In this work, we consider an extension of the MSSM with vector-like supermultiplets  
to increase the Higgs boson mass~\cite{Moroi:1991mg,Babu:2004xg,Babu:2008ge,Martin:2009bg, Asano:2011zt},
and investigate how much the Higgs mass can be raised while explaining the deviation of the muon $g-2$.
As for mediation mechanisms, gauge-mediated SUSY breaking (GMSB) models and gravity-mediated SUSY breaking models are considered. 

The rest of this paper is organized as follows. In Sec.~\ref{sec:setup}, we introduce the framework.
A generic discussion on the Higgs mass and the muon $g-2$ is given in Sec.~\ref{sec:mh_and_g-2}.
In Sec.~\ref{sec:RGE}, we discuss the renormalization group evolutions of the model parameters
with two-loop $\beta$ functions.
The results of numerical analyses are shown in Sec.~\ref{sec:GMSB} for GMSB, and Sec.~\ref{sec:mSUGRA} for gravity-mediated breaking. In GMSB models, the Higgs mass can be larger than 135 GeV 
(130 GeV) in the region where the deviation of the muon $g-2$ is explained at the $2\sigma$ ($1\sigma$) level.
In gravity mediation case with universal soft masses, 
the Higgs mass can be as large as about 130 GeV in the region where the muon $g-2$ is explained.
Interestingly, it overlaps with the coannihilation region.
Sec.~\ref{sec:summary} is devoted to summary and discussion.
In Appendix A, one-loop corrections to the Higgs potential, generated by the extra-matters, are shown, and in Appendix B, 
renormalization group equations at the two-loop level are shown.

\section{Setup}
\label{sec:setup}

We consider models consisting of
 the MSSM particles and a vector-like pair of complete SU(5) multiplets, ${\bf 10}=(Q', U', E')$ and 
$\overline{\bf 10}=(\bar{Q}', \bar{U}', \bar{E}')$. The extra 
matters form the Lagrangian,
\begin{eqnarray}
W = Y' Q' H_u U' + Y'' \bar{Q}' H_d \bar{U}' + M_{Q'} Q' \bar{Q}' + M_{U'} U' \bar{U}' + 
M_{E'} E' \bar{E}',
\end{eqnarray}
and the SUSY breaking terms,
\begin{eqnarray}
-\mathcal{L}_{\rm soft} &=& m_{Q'}^2 |\tilde{Q}'|^2 + m_{\bar{Q}'}^2 |\tilde{\bar{Q}'}| + m_{U'}^2 |\tilde{U}'|^2 + m_{\bar{U}'}^2 |\tilde{\bar{U}'}| + m_{E'}^2 |\tilde{E}'|^2 + m_{\bar{E}'}^2 |\tilde{\bar{E}'}|  \nn
&+& B_{Q'} M_{Q'} \tilde{Q}' \tilde{\bar{Q}'} + B_{U'} M_{U'} \tilde{U}' \tilde{\bar{U}'} + B_{E'} M_{E'} \tilde{E}' \tilde{\bar{E}'} \\
&+& A' Y' \tilde{Q}' H_u \tilde{U}' + A'' Y'' \tilde{\bar{Q}'} H_d \tilde{\bar{U}'} + h.c. ,
\end{eqnarray}
where the tilde represents the scalar component of the chiral superfield. Here, $m_{Q', U', E'}^2$
are soft scalar masses, $B_{Q', U', E'}$ are the B-terms, and $A' Y'$ and $A'' Y''$ are trilinear scalar couplings. 
In the models, it is supposed that a parity is assigned to the extra vector-like matter to avoid large mixings 
with the MSSM particles, though this parity is considered to be violated weakly 
in order to avoid cosmological difficulties due to stable exotic particles.

There are two Yukawa interactions with $Y'$ and $Y''$ in the superpotential. The latter term couples to 
the down-type Higgs. Since it reduces the lightest Higgs boson mass especially when $\tan\beta$ is 
large (see discussion in Sec.~\ref{sec:mh_and_g-2}), we assume $Y'' \ll Y'$ in the following.

The (SUSY-invariant) vector-like masses $M_{Q', U', E'}$ determine the mass scale of the extra 
multiplets and are assumed to be in the TeV scale. Although there is no {\it a priori} mechanism to 
explain the scale, 
the vector-like masses may be generated dynamically as well as the $\mu$-term of the MSSM
as in the NMSSM.
Very recently, it has been shown that a non-anomalous discrete $R$-symmetry 
can naturally explain the existence of such TeV scale vector-like matters~\cite{Asano:2011zt}.
In this work, we consider the SUSY invariant masses of the vector-like matters as parameters for generality. 

\section{Higgs mass and muon $g-2$}
\label{sec:mh_and_g-2}

The extra matters couple to the up-type Higgs, which behaves as an SM-like Higgs when 
the heavy Higgs bosons decouple. The scalar potential of the lightest Higgs boson receives corrections
from the extra matters, similarly to the top (s)quark in the MSSM. 
Assuming $Y''\ll Y'$ and large $\tan\beta$, 
it is approximately given by~\cite{Babu:2008ge,Martin:2009bg}
\begin{eqnarray}
  \Delta m_h^2 \simeq 
  \frac{3 Y'^4 v^2}{4\pi^2} \left[
  \ln \frac{m_S^2}{m_F^2} - 
  \frac{1}{6}\left(1-\frac{m_F^2}{m_S^2}\right)\left(5-\frac{m_F^2}{m_S^2}\right) + 
  \frac{A'^2}{m_S^2}\left(1-\frac{m_F^2}{3m_S^2}\right) - \frac{1}{12} \frac{A'^4}{m_S^4}
  \right],
  \label{eq:higgs_mass}
\end{eqnarray}
from the one-loop effective potential of the scalar Higgs. Here, $m_S$ is a mass of the scalar 
vector-like matter, which differs from the mass of the fermionic one, $m_F$, by the soft parameters, 
$m_S^2 = m_F^2 + m_{\rm soft}^2$. The corrections are proportional to $Y'^4$, and the first term 
in the bracket is enhanced when there is a hierarchy between $m_S$ and $m_F$, while the terms 
dependent on the trilinear coupling $A'$ become effective when $A'$ is properly large. 
It has been studied,  e.g. in \cite{Martin:2009bg}, that,  by maximizing the latter 
contribution (`the maximal mixing scenario'), the correction can be as large as $\Delta m_h 
\sim 20 - 50$ GeV  for $Y' \simeq 1$ compared to the MSSM value. 
Note that, however, this situation is usually not 
realized if the renormalization group evolution (RGE) is considered due to 
the infrared fixed-point behaviour~\cite{Martin:2009bg} (see Sec.~\ref{sec:RGE}).

The lightest Higgs boson mass is raised by the top (s)quarks within the MSSM. 
It can be shown (e.g., by using the FeynHiggs package \cite{FeynHiggs2.7})
that the mass can reach 130 GeV if the top squarks are as heavy as $2-3$ TeV
and the scalar top trilinear coupling, $A_t$, is large. 
However, again the RGE significantly affects the soft parameters,
particularly when $Y'$ is large, as will be discussed in Sec.~\ref{sec:RGE}.

It should be commented that, if $Y''$ is large, the lightest Higgs boson mass receives a sizable 
negative contribution, $\Delta m_h^2 \sim -\frac{3v^2}{48\pi^2} Y''^4 \frac{\mu^4}{m_S^4}$. This 
term corresponds to the last term in (\ref{eq:higgs_mass}) with $A'Y'  \to \mu Y''$, while the other 
terms are suppressed by powers of $\tan\beta$ when $\bar{U}'$ contributes to the lightest Higgs 
potential. Thus, $Y' \gg Y''$ is required to enhance the lightest Higgs boson mass.

The discrepancy between the experimental and SM values of the muon $g-2$, $\Delta a_\mu \equiv 
a_\mu({\rm exp}) - a_\mu({\rm SM}) = (26.1 \pm 8.0) \times 10^{-10}$\cite{g-2_hagiwara}, is easily saturated in the SUSY 
models when $\tan\beta$ is large. The SUSY contributions consist of chargino ($\chi^\pm$) 
and neutralino ($\chi^0$) ones~\cite{Moroi:1995yh}:
\begin{eqnarray}
  \Delta a_\mu(\chi^\pm) &\simeq& 
  \frac{\alpha_2 m_\mu^2}{8\pi m_{\rm soft}^2} {\rm sgn}(\mu M_2) \tan\beta, \\
  \Delta a_\mu(\chi^0) &\simeq& 
  \frac{\alpha_Y m_\mu^2}{24\pi m_{\rm soft}^2} {\rm sgn}(\mu M_1) \tan\beta + \cdots,
\end{eqnarray}
where $m_{\rm soft}$ represents the soft parameters and the Higgsino mass $\mu$.
It is noticed that, since the SUSY contributions are proportional to $\tan\beta$, 
they can be enhanced for large $\tan\beta$, and $\Delta a_\mu$ can be 
as large as $O(10^{-9})$ for $\tan\beta = O(10)$. Also, the positive discrepancy of the 
muon $g-2$ prefers a positive ${\rm sgn}(\mu M_{1,2})$ in most of the parameter space.

The chargino contribution dominates the SUSY contributions when all the SUSY (soft) masses 
are almost the same. The SUSY mass $m_{\rm soft}$ of the chargino contribution depends on the 
mass of the left-handed smuon (the muon sneutrino), the Wino mass and $\mu$. Thus, it decreases 
especially when $\mu$ becomes large. 
When the $\mu$ term is large, the neutralino contribution becomes effective, where the lightest 
neutralino is purely Bino, and both the left- and right-handed smuons contribute. 
In this case, $m_{\rm soft}$ is insensitive to $\mu$. It can be checked that this 
contribution decreases when the Bino and/or the smuons become heavy. 
It is also mentioned that the SUSY contributions are independent of the trilinear coupling of the 
smuon.

We use SuSpect package~\cite{suspect2} for calculating the sparticle spectrum, 
which is modified to include two-loop renormalization group running effects from the extra matters 
(cf. App.~\ref{app:RGE}).
We evaluate the lightest Higgs boson mass at the two-loop level for the MSSM 
contribution by using the FeynHiggs package \cite{FeynHiggs2.7},
and the contribution of the extra vector-like multiplet is evaluated at the one-loop level 
by using the formula in App. \ref{app:mh}.\footnote{
The renormalization scale, $Q$, is set to be the geometrical average of the stop masses in 
the analysis. Since the extra matters have a large Yukawa coupling, the theoretical uncertainty 
of the Higgs potential due to the extra matters is considered to be large. Although it can be reduced 
by evaluating the potential at the two-loop level, the calculation is beyond the scope of this 
paper.}
The muon $g-2$ is estimated by FeynHiggs. Note that the extra matters do 
not contribute to muon $g-2$ at the one-loop level.
Top quark mass is taken to be $m_t=173.1$ GeV.

\section{Renormalization Group Evolutions}
\label{sec:RGE}

The extra vector-like matters affect the renormalization group evolution of the MSSM parameters. 
The perturbative gauge coupling unification is preserved, and particularly the SU(3) gauge coupling 
constant remains large up to the GUT scale if the vector-like matter ${\bf 10}+\overline{\bf 10}$ exists at the TeV scale,
because the $\beta$ function vanishes at the one-loop level (see \cite{Martin:2009bg} for the two- and 
three-loop running). Furthermore, when the extra Yukawa couplings are large, the model parameters 
which are relevant for the Higgs boson mass, $Y'$, $A'$, and $A_t$, 
are likely to flow to an infrared fixed point, as discussed below.
In particular, it turns out that the 
maximal mixing scenario (i.e., maximizing the contribution of the 
last two terms in Eq.~(\ref{eq:higgs_mass}))
is difficult to be realized unless the SUSY 
breaking is mediated at a low scale. In this section, the RG behaviours are studied, 
particularly focusing on the Higgs boson mass and the muon $g-2$. 
In the numerical analysis, we have used the two-loop RG equations listed in App.~\ref{app:RGE}.\footnote{Although large coupling constants at the GUT scale may enhance threshold corrections 
of the GUT breaking, 
the corrections depend on  details of the GUT structure, and are neglected in the analysis for simplicity. }

The gaugino mass behaviours and their contributions to the other soft parameters are much 
different from the MSSM case \cite{Martin:2009bg}. Since the gauge couplings are large in high 
scale, the gauginos are heavy at the scale. It is emphasized that the two-loop $\beta$ function is 
crucial for the gluino mass, which reduces the running gluino mass about 40\% at the weak scale compared to 
the input at the GUT scale, though it does not evolve at the one-loop level.

The soft parameters to which the up-type Higgs and the extra matters do not directly couple are 
strongly raised by the 
gaugino masses because both the gauge coupling constants and the gaugino masses are large in 
high scale. This becomes prominent if the mediation scale is higher. In particular, the scalar lepton 
masses grow more rapidly at low energy than the case of usual MSSM.
Thus, high scale mediation models tend to suppress the SUSY contributions to the muon $g-2$ 
for a fixed Higgs boson mass compared to low scale mediations.

 \begin{figure}[t]
\begin{center}
\includegraphics[scale=0.69]{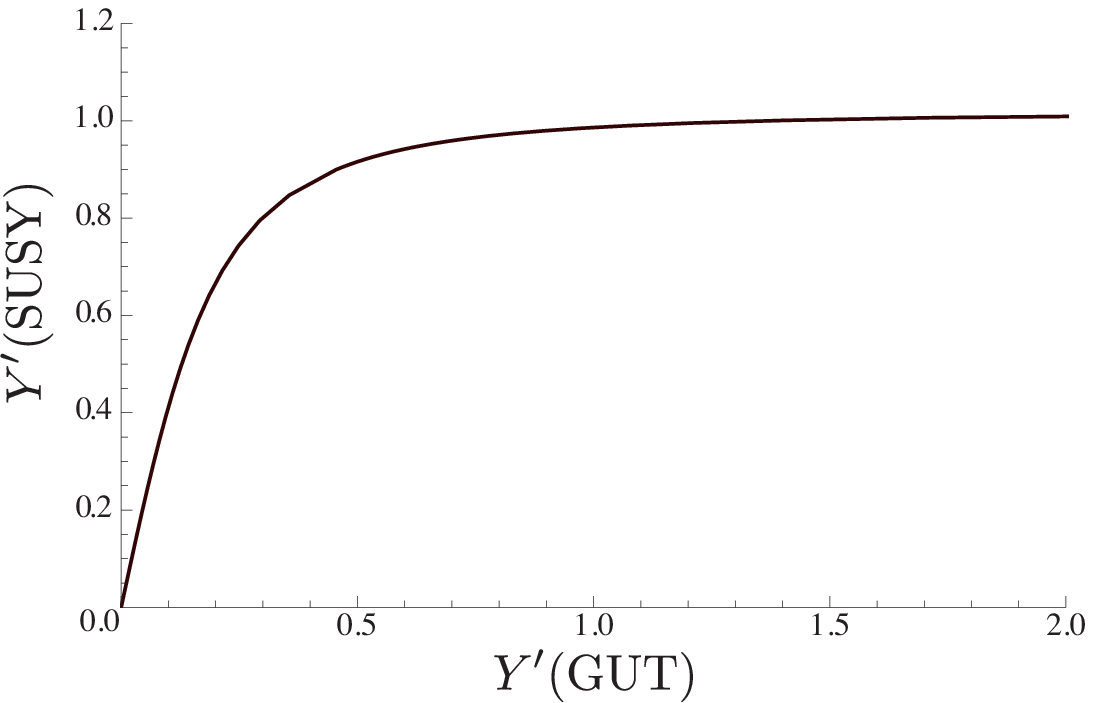}
\hspace*{3mm}
\includegraphics[scale=0.69]{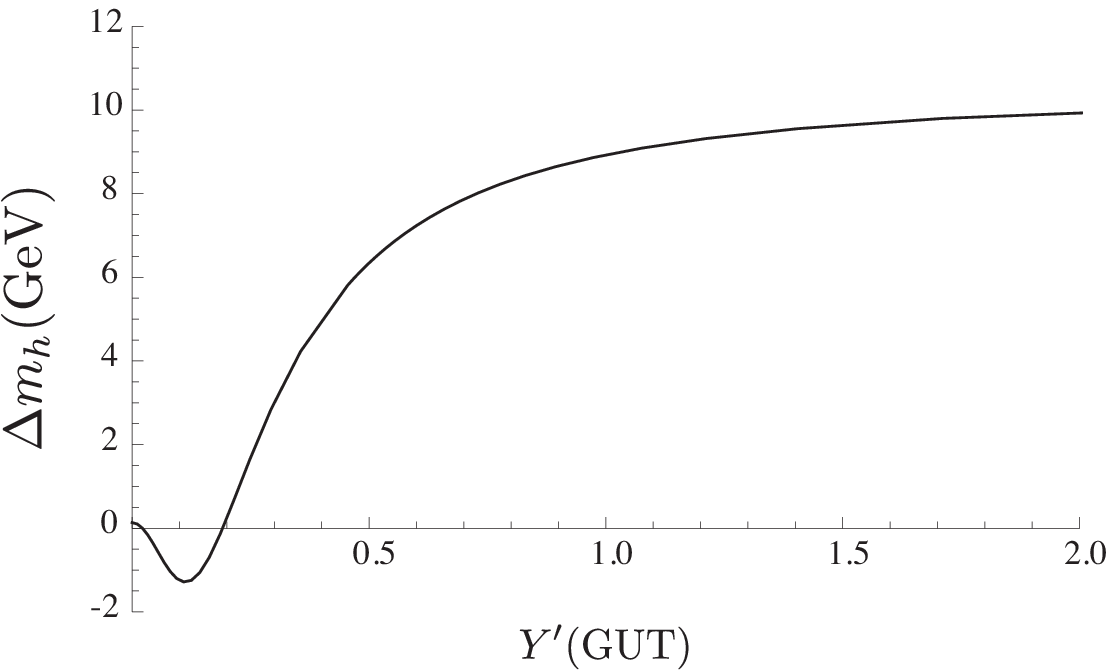}
\caption{$Y'({\rm SUSY})$ (left) and $\Delta m_h$ (right) as a function of $Y'({\rm GUT})$, 
where the gaugino mass is set $M_{1/2} = 1.5$ TeV and the others zero at the GUT scale. The 
vector mass of the extra matter is taken to be 600 GeV at the weak scale, and the Higgs boson 
mass is assumed to be 120 GeV within the MSSM.}
\label{fig:Y10}
\end{center}
\end{figure}

The Yukawa coupling of the vector-like multiplet is relevant for the lightest Higgs boson mass. 
Since the $\beta$  function of the strong gauge coupling vanishes at the one-loop level, the Yukawa coupling flows to the 
infrared fixed point~\cite{Martin:2009bg}, as is noticed from $\beta$ function of $Y'$ in App.~\ref{app:RGE} [Eq.~(\ref{eq:betaY})].  The Yukawa coupling constant at the weak scale is plotted against the value 
at the GUT scale in Fig.~\ref{fig:Y10}. It is noticed that the coupling becomes as large as $Y' \simeq 1$ 
at the weak scale if it is larger than $\sim 0.5$ at the GUT scale. Thus, the Higgs boson mass can be 
raised by $O(10)$ GeV for $Y' \gsim 0.5-1$ at the GUT scale, depending on the soft scalar masses of the 
vector-like multiplets. On the other hand, $Y'$ cannot be much larger than unity at the weak scale 
even if it is much larger at high scale.

 \begin{figure}[t]
\begin{center}
\includegraphics[scale=0.7]{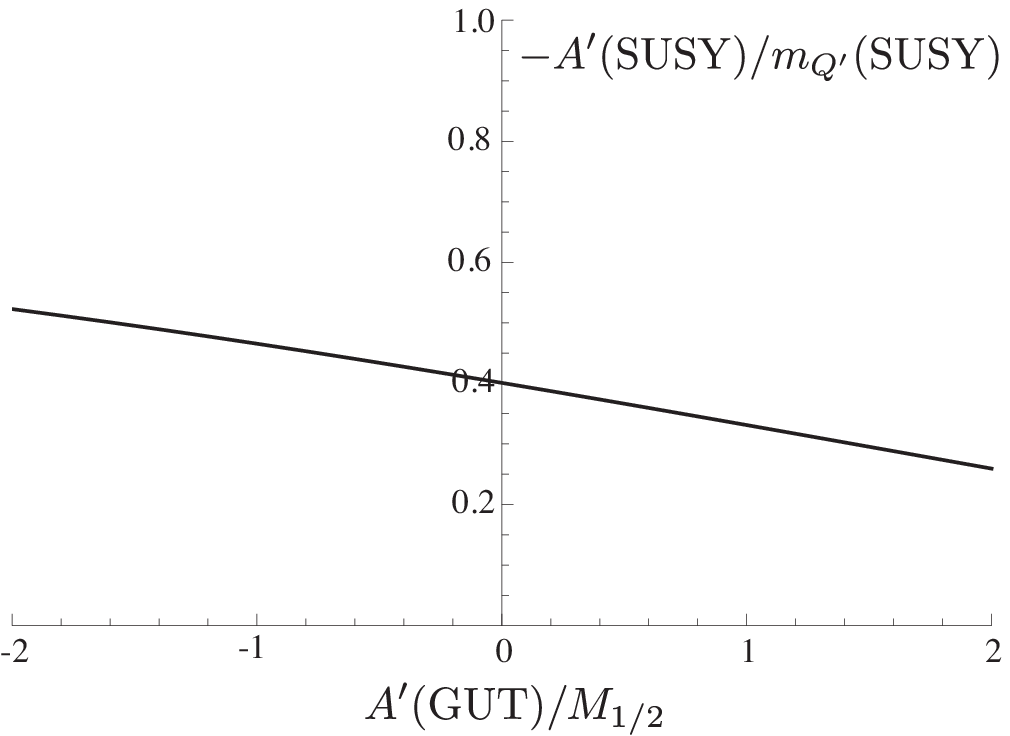}
\hspace*{3mm}
\includegraphics[scale=0.7]{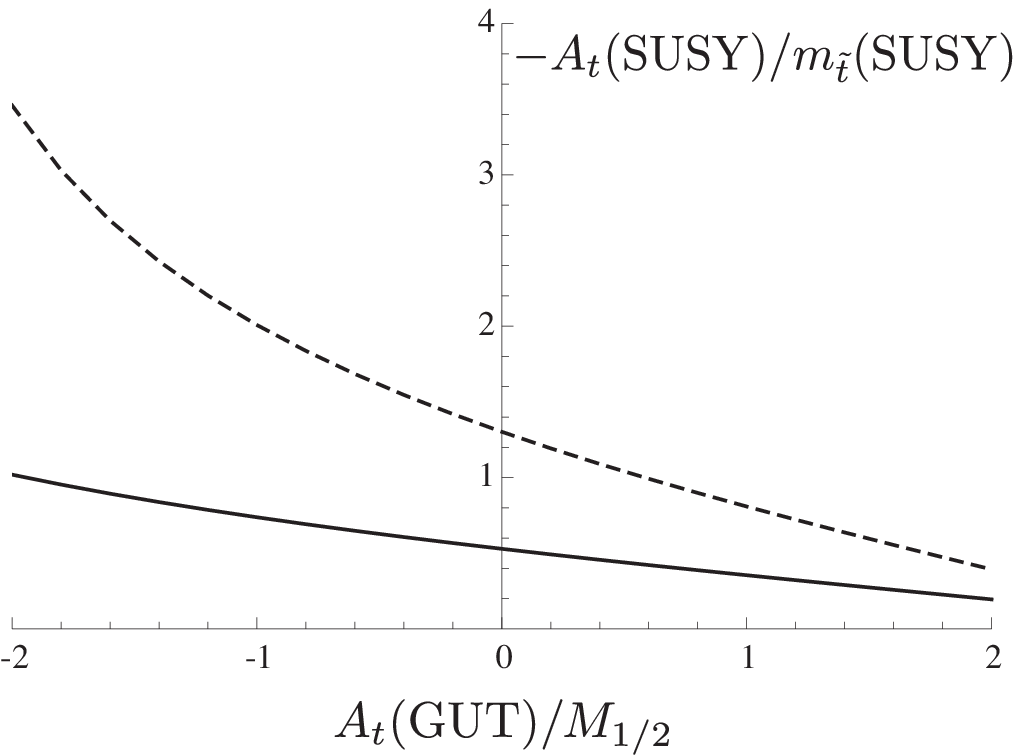}
\caption{$A'({\rm GUT})/M_{1/2}$ vs $-A'({\rm SUSY})/m_{Q'}({\rm SUSY})$ (left) and 
$A_t({\rm GUT})/M_{1/2}$ vs $-A_t({\rm SUSY})/m_{\tilde t}({\rm SUSY})$ (right), where 
the Yukawa coupling is $Y' = 3$ (solid) and $Y' = 0$ (dashed) at the GUT scale. 
The gaugino mass is set $M_{1/2} = 1.5$ TeV and the others zero at the GUT scale.}
\label{fig:A10AT}
\end{center}
\end{figure}

A large trilinear coupling of the vector-like matter, $A'$, at the weak scale can enhance the Higgs 
boson mass if the maximal mixing scenario is realized in Eq.~(\ref{eq:higgs_mass}). 
Since the Higgs couples to the extra vector-like matters, the $A'$ parameter is 
also affected by the Yukawa interactions during the 
RG evolution. As is noticed from the $\beta$-function of $A'$ in App.~\ref{app:RGE} [Eq.~(\ref{eq:betaA})], 
this coupling $A'$ is also likely to be focused to an infrared fixed point, 
which is estimated to be $\sim 0.5 M_{1/2}$~\cite{Martin:2009bg}, when $Y'$ is large. Consequently, $A'/m_{Q'}$, which determines the correction to the 
Higgs boson mass, becomes rather insensitive to $A'$ at the GUT scale, as shown in Fig.~\ref{fig:A10AT}.
In particular, since the ratio turns out to be $\sim 0.4$, 
it is difficult to realize the maximal mixing scenario, which requires $A'/m_{Q'}\sim \sqrt{6}$ for $m_{Q'}\gg M_{Q'}$.
In fact, it is found that the correction to the Higgs boson mass from the extra matter changes only about 1 GeV 
even if $A'$ is varied from $-2 M_{1/2}$ to $+2 M_{1/2}$ at the GUT scale (cf. Fig.~\ref{fig:A_M12}). 
Therefore, in the present setup, the Higgs mass is likely to be enhanced mainly by 
a large hierarchy between the scalar and fermion masses of the extra vector-like matter.

The $\beta$ function of the scalar top trilinear coupling also depends on the extra Yukawa coupling 
accompanied with $A'$ (see Eq.~(\ref{eq:betaAtop}) 
of App.~\ref{app:RGE}). The contribution suppresses $A_t$ during the 
RG running down to the weak scale. In Fig.~\ref{fig:A10AT}, $A_t/m_{\tilde t}$, which determines 
the $A_t$ contribution to the Higgs mass, is plotted as a function of $A_t/M_{1/2}$ at the GUT scale. 
It is found that $A_t$ is suppressed significantly if $Y'$ is large (the solid line in the figure) compared to 
the result with $Y'=0$ (the dashed line). Consequently, the $A_t$ at the weak scale becomes
rather insensitive to 
input values at the GUT scale, and the correction to the Higgs boson mass results in less than 1 GeV for 
$-M_{1/2} \leq A_t({\rm GUT}) \leq M_{1/2}$. In particular, it is difficult to maximize the correction to the Higgs 
boson mass by tuning $A_t$ within the MSSM. 

Let us also mention the RG evolution of the $m_{H_u}^2$ and its effect on 
the $\mu$ parameter. In addition to the top squarks, 
the extra vector-like multiplets contribute at the one-loop level and draw down the $m_{H_u}^2$ 
significantly during the RG running when $Y'$ is large (cf. Eq.~(\ref{eq:betaHu})),
which results in a large $\mu$ parameter.

The RG behaviours of the parameters which are relevant for the Higgs boson mass indicate that 
the correction to the Higgs boson mass is enhanced when the scalar masses are hierarchical 
compared to the fermion masses in the chiral multiplets rather than enlarging the trilinear couplings. 
On the other hand, a large soft SUSY breaking scale generally reduces the SUSY contribution to the muon $g-2$. 
Therefore, it is nontrivial whether a large enhancement of the Higgs mass
and an explanation of the muon $g-2$ are simultaneously realized.
In the next sections, 
the Higgs boson mass and the muon $g-2$ are studied for some typical boundary 
conditions at high scale. 

Before closing this section, let us comment on another Yukawa coupling, $Y''$. The RG evolution of $Y''$ 
is similar to that of $Y'$, though the $\beta$ function is omitted in App.~\ref{app:RGE}. Similarly to 
Fig.~\ref{fig:Y10}, $Y''$ has an infrared fixed point close to 1. As was explained in the previous 
section, the large $Y''$ coupling reduces the Higgs boson mass. The reduction becomes prominent when 
both of the Yukawa couplings are large. In fact, $\Delta m_h$ is smaller by $\sim 1-10$ GeV for $Y'= Y'' 
= 3$ at the GUT scale than that for $Y'' = 0$ with $Y'({\rm GUT})=3$,
depending on the mass spectrum.
Thus, $Y''$ is required to be tiny at the 
GUT scale in order to enhance the correction to the Higgs boson mass. Such a tiny $Y''$ may be 
realized by assigning an extra charge on $H_d$ such as the Peccei-Quinn charge. It might be also 
interesting if the smallness is related to a solution to the $\mu$ problem.
In this paper, $Y''\ll 1$ is assumed, 
and we neglect the contribution of $Y''$ in the following analysis.

 \begin{figure}[t]
\begin{center}
\includegraphics[width=10cm]{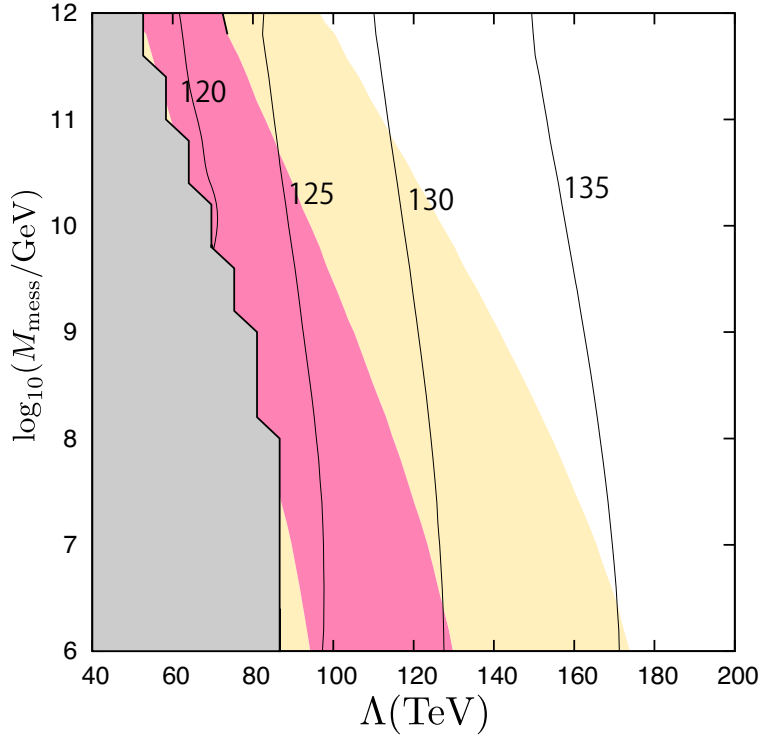}
\caption{Higgs mass and muon $g-2$ in GMSB models with vector-like matters
in $(\Lambda, M_{\rm mess})$ plane, for $\tan\beta=30$ and $M_{Q',U',E'}=600$ GeV.
The stau becomes tachyonic when $\Lambda$ is small (gray shaded region).
The red (yellow) shaded region can explain the muon $g-2$ anomaly
within 1$\sigma$ (2$\sigma$).
The vertical solid lines are contours of the Higgs mass.
 }
\label{fig:GMSB_tanb30}
\end{center}
\end{figure}

 \begin{figure}[t]
\begin{center}
\includegraphics[width=10cm]{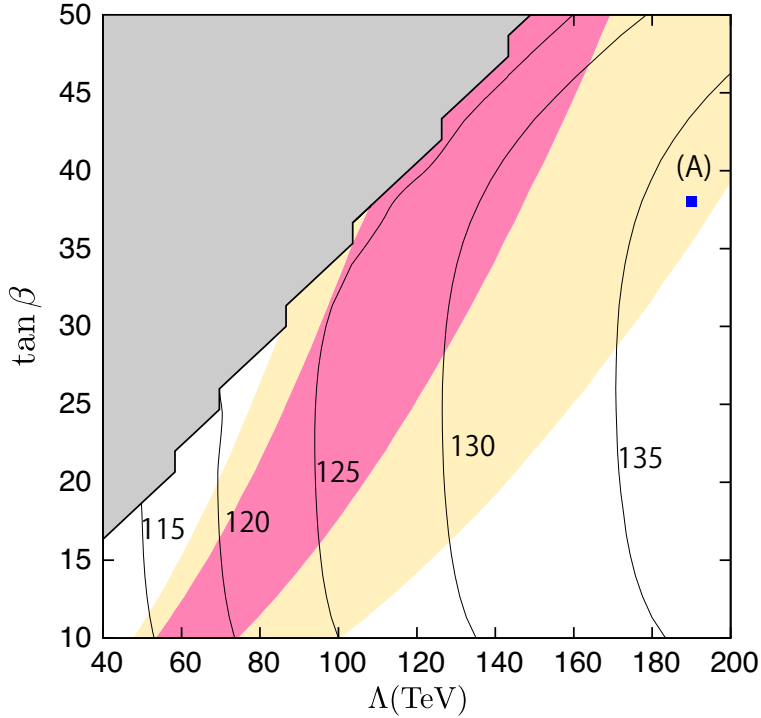}
\caption{Higgs mass and muon $g-2$ in GMSB models with vector-like matters
in $(\Lambda, \tan\beta)$ plane, for $M_{\rm mess}=5\times 10^5$ GeV and $M_{Q',U',E'}=600$ GeV.
The stau becomes tachyonic when $\Lambda$ is small (gray shaded region).
The red (yellow) shaded region can explain the muon $g-2$ anomaly
within $1\sigma$ ($2\sigma$).
The vertical solid lines are contours of the Higgs mass.
The blue square corresponds to the model point (A) in Table.~\ref{tab:GMSB}.
 }
\label{fig:GMSB_mess5d5}
\end{center}
\end{figure}

\begin{table}[t]
\begin{center}
\begin{tabular}{|c|c|c|c|c|c|c|}
\hline
& $\Lambda$ & $M_{\rm mess}$ & $\tan\beta$ & $M_{Q',U',E'}$
& $\Delta a_\mu$
& $m_h$ ($m_h^{\rm MSSM}$) 
\\ \hline
(A) 
& $1.9\times 10^5$ & $5\times 10^5$ & 38 & 600
& $10.9 \times 10^{-10}$ 
& 136 (118) 
\\ \hline
(B) 
& $ 3\times 10^5$ & $1\times 10^6$ & 40 & 600
& -- 
& 143 (119) 
\\ \hline
\end{tabular}
\vspace{1em}

\begin{tabular}{|c|c|c|c|c|c|c|c|c|}
\hline
& $m_{\tilde{g}}$ 
& $m_{\tilde{t}_1}$ & $m_{\tilde{T'}_{1-4}}$ 
& $\tilde{\mu}_{\rm L}$ & $\tilde{\mu}_{\rm R}$ & $\tilde{\tau}_1$ 
& $\chi^{\pm 1}_1$ 
& $\chi^0_1$ 
\\
\hline
(A) 
& 1758 & 2460 & 2418--2835 
& 764 & 353 & 264 
& 545 & 273 
\\ \hline
(B) 
& 2675 & 3780 & 3647--4290
& 1212 & 567 & 471
& 854 & 431
\\ \hline
\end{tabular}
\caption{A part of the mass spectrum and the value of deviation of the muon $g-2$,
for a GMSB model with vector-like fields, 
for a model point (A) in Fig.~\ref{fig:GMSB_mess5d5}.
$m_{\tilde{T'}_{1-4}}$ represents the scalar masses of
vector-like squarks.
For comparison we have also shown the 
Higgs mass which would be obtained without the vector fields, as $m_h^{\rm MSSM}$.
All the masses are written in units of GeV.
For illustration, we also show another model point (B) with a Higgs mass above 140 GeV. 
The neutralino contribution to the muon $g-2$ is comparable to that of the chargino at (A).
}
\label{tab:GMSB}
\end{center}
\end{table}

 \begin{figure}[t]
\begin{center}
\includegraphics[width=10cm]{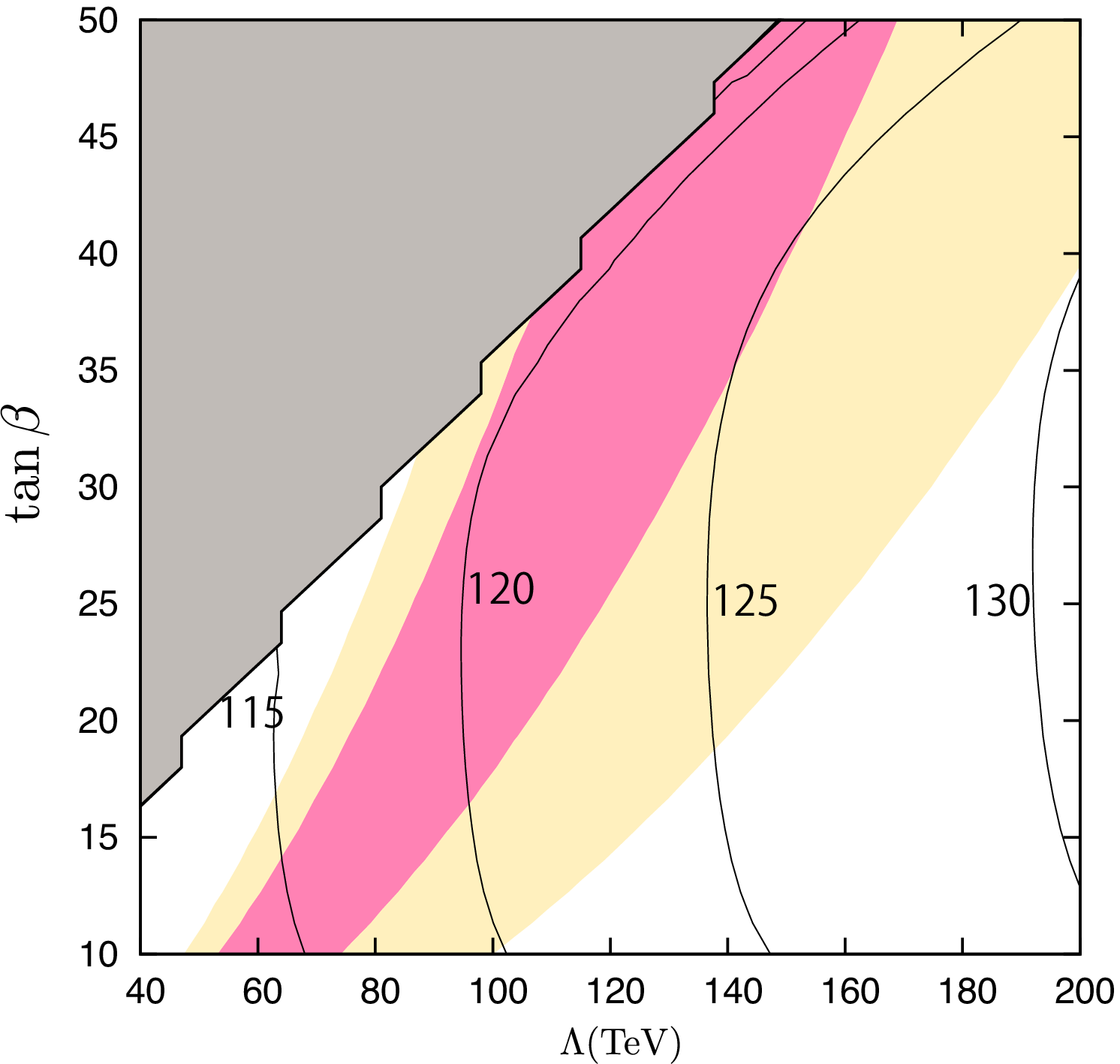}
\caption{The same as Fig.~\ref{fig:GMSB_mess5d5} but for $M_{Q',U',E'}=1$ TeV.}
\label{fig:GMSB_mess5d5_1TeV}
\end{center}
\end{figure}

\section{Higgs mass and muon $g-2$ in GMSB models with vector-like matters}
\label{sec:GMSB}
In this section, we investigate the Higgs mass and the muon $g-2$ in the framework 
of the GMSB models~\cite{Giudice:1998bp} with 
the extra vector-like matters.
The simplest GMSB model is parametrized by 5 parameters:
$M_{\rm mess}$ (messenger mass scale), 
$\Lambda = F_{\rm mess}/M_{\rm mess}$ 
(the ratio of the non-holomorphic mass squared to the holomorphic mass for messengers),
$N_5$ (messenger number), 
$\tan\beta$, and ${\rm sgn}(\mu)$. 
The gravitino mass is irrelevant to our analysis.
In the analysis, the sign of $\mu$ is fixed to be ${\rm sgn}(\mu) = 1$ to explain the anomaly 
of the muon $g-2$. On the other hand, the messenger number is set $N_5=1$. 
Note that, since the vector-like matters, ${\bf 10} + \overline{\bf 10}$, at the TeV scale
significantly affect the $\beta$ functions of the gauge couplings as emphasized in Sec.~\ref{sec:RGE}, 
if the messenger number is $N_5 \ge 2$, the running gauge coupling can be non-perturbative 
below the GUT scale unless the messenger scale is high.

The soft SUSY breaking parameters are induced at the messenger scale by the GMSB mechanism 
for the new vector fields as well as the MSSM. The (SUSY-invariant) vector masses are
 $M_{Q',U',E'}=600$ GeV, if not otherwise specified.\footnote{With light vector-like fermions,
the Higgs production cross section through gluon fusion at the LHC is  reduced,
but only by a few percent for $M_{Q',U'}=600$ GeV~\cite{Ishiwata:2011hr}.
The corrections to the electroweak precision observables are also small~\cite{Martin:2009bg}.
The bound from the direct searches at the LHC~\cite{CMS:EXO-11-051} is also evaded for
$M_{Q',U'}=600$ GeV.
}
 The $B$ parameters of the vector-like matters, 
$B_{Q', U', E'}$, are set to be 0 at the messenger scale, for simplicity. They are related to the 
mechanism which generates $M_{Q',U',E'}$, though the GMSB mechanism does not induce them.
Therefore, there are essentially 3 parameters in our setup: $M_{\rm mess}$, $\Lambda = F_{\rm mess}/M_{\rm mess}$, 
$\tan\beta$, in addition to the SUSY invariant masses for the vector-like matters, $M_{Q',U',E'}$, and
the Yukawa coupling $Y'$.

In Fig.~\ref{fig:GMSB_tanb30}, we show contours of the Higgs mass and the muon $g-2$ in $(\Lambda, M_{\rm mess})$ plane for $\tan\beta=30$.
Here and hereafter, the Yukawa coupling of the vector-like matter is taken to be $Y'=1$, for simplicity.
As discussed in Sec.~\ref{sec:RGE}, this is close to the fixed point value, 
and it cannot be larger to maintain the perturbativity up to the GUT scale. 
It is found that the Higgs mass is enhanced for larger $\Lambda$, because the SUSY particles are 
heavier, including the stops and the new vector fields. However, larger slepton masses suppress 
the SUSY contribution to the muon $g-2$. 
On the other hand, when the messenger scale increases, the  SUSY particle masses, particularly 
the slepton mass and $\mu$, become larger, which reduces the contribution to the muon 
$g-2$, whereas the Higgs mass is less sensitive to the messenger scale because the SUSY breaking scale 
appears through the logarithm in the correction to the Higgs boson mass when the scalar is much 
heavier than its fermionic partner. Thus, lower-scale GMSB models are favored to enhance the 
Higgs boson mass in the light of the muon $g-2$ anomaly.

The contours of the Higgs mass and the muon $g-2$ are displayed in $(\Lambda, \tan\beta)$ 
plane for $M_{\rm mess}=5\times 10^5$ GeV in Figs.~\ref{fig:GMSB_mess5d5}.
It is noteworthy that the Higgs mass can be larger than 130 GeV (135 GeV)
in the region where the muon $g-2$ agrees with the experimental value 
within the 2$\sigma$ (1$\sigma$) uncertainty.
This is one of the main conclusions of this paper.
As a reference point, 
a part of the mass spectrum of the SUSY particles, as well as the value of the deviation of the muon $(g-2)$,
for $\Lambda = 190$~TeV, $\tan\beta=38$ is listed in Table.~\ref{tab:GMSB}.

It should be emphasized that the Higgs mass becomes larger than 130 GeV in wide parameter regions 
by the additional vector-like multiplet. 
This is contrasted to the normal GMSB models, where it is difficult to realize the Higgs boson mass 
of 130 GeV.
It is also interesting that the Higgs boson mass can exceed 140 GeV if the $g-2$ 
constraint would not be imposed, as shown in Table.~\ref{tab:GMSB}.

So far, it was assumed that the SUSY invariant mass of the vector field is relatively small, 
$M_{Q',U',E'}=600$ GeV. 
To see the dependence on the vector mass, the result for $M_{Q',U',E'}=1$ TeV is shown
in Fig.~\ref{fig:GMSB_mess5d5_1TeV}. 
Since a large SUSY invariant mass decreases the hierarchy between the scalar and fermion 
fields of the multiplet, the vector-matter contributions to the Higgs mass are suppressed.
In the region where the muon $g-2$ is consistent with the experimental value at the 2$\sigma$, 
the maximal Higgs mass becomes lower than 130 GeV.

\section{Higgs mass and muon $g-2$ in mSUGRA models with vector-like matters}
\label{sec:mSUGRA}

The Higgs mass and the muon $g-2$ are studied in the mSUGRA models in the presence of the 
extra vector-like multiplet. At the GUT scale, the boundary conditions are specified by $M_{1/2}$, 
$m_0 \equiv m_0({\rm MSSM}) = m_{Q'} = m_{\bar Q'} = m_{U'} = m_{\bar U'} = m_{E'} 
= m_{\bar E'}$, $A_0 = A_0({\rm MSSM}) = A'$, $\tan\beta$, and ${\rm sgn}(\mu)$. Here, 
$m_0({\rm MSSM})$ and $A_0({\rm MSSM})$ represent the universal scalar mass and trilinear 
coupling in the MSSM, respectively. On the other hand, the SUSY invariant mass $M_{Q', U', E'}$ 
and the $B$ parameters $B_{Q', U', E'}$ are set at the weak, since they are independent of the 
RG evolutions of the other parameters. In particular, we can set $B_{Q'}=B_{U'}=B_{E'}=0$ without 
modifying the following result. It can be checked that they are almost irrelevant for the Higgs mass 
and the muon $g-2$, though they appear in the scalar mass matrix of the vector matter.

\begin{figure}[t]
\begin{center}
\includegraphics[width=10cm]{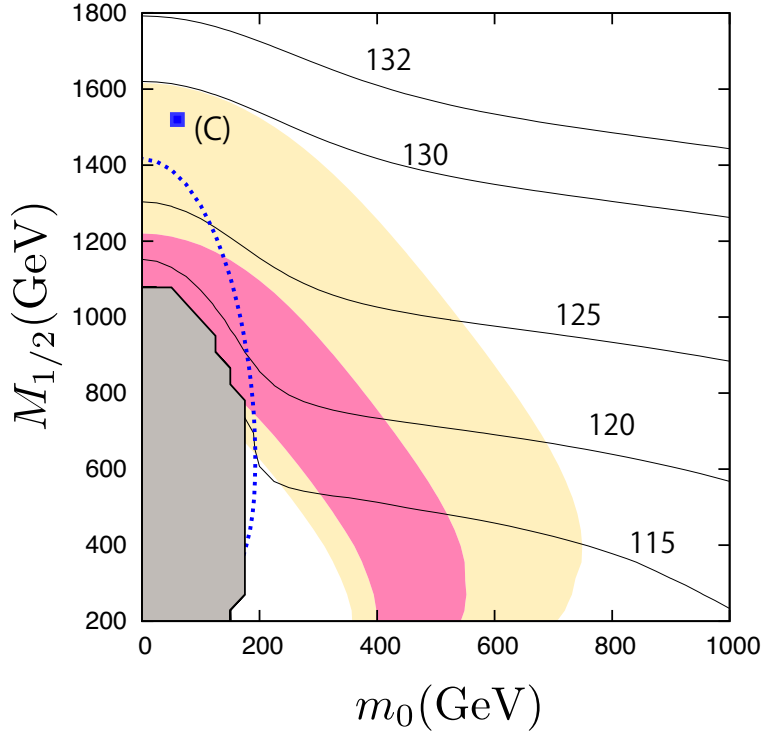}
\caption{Higgs mass and muon $g-2$ in mSUGRA models with vector-like matters
in $(m_0, M_{1/2})$ plane, for $\tan\beta=40$, $A_0=0$, 
$M_{Q',U',E'}=600$ GeV.
The blue square corresponds to the model point (C) in Table.~\ref{tab:mSUGRA}.
The stau is tachyonic in the gray region.
}
\label{fig:mSUGRA_tb40_A0_0}
\end{center}
\end{figure}

In Fig.~\ref{fig:mSUGRA_tb40_A0_0}, 
we show the Higgs mass and the muon $g-2$ in $(m_0, M_{1/2})$ plane, for $A_0=0$, $\tan\beta=40$,  $M_{Q',U',E'}=600$ GeV, and $Y'=1$. 
When $m_0$ and/or $M_{1/2}$ becomes larger, the scalar masses increases, and the SUSY 
contribution to the muon $g-2$ is suppressed. On the other hand, the contribution to the Higgs 
boson mass is less sensitive to $m_0$, because the RG evolution of the colored SUSY particles 
is dominantly controlled by the gluino mass. 
It is found that in the region where the SUSY contribution is consistent with the muon $g-2$ at the 
$2\sigma$ level, the maximal value of the Higgs mass can be as large as 130 GeV in Fig.~\ref{fig:mSUGRA_tb40_A0_0}. 

The parameter region where the Higgs mass is large and the muon $g-2$ is consistent with the 
experiment is interesting from the cosmological view points. 
In this region, $m_0=0-200$ GeV and $M_{1/2}\sim 1400$ GeV, the masses of the neutralino 
becomes close to that of the lightest stau.
In Fig.~\ref{fig:mSUGRA_tb40_A0_0}, we draw the (blue dotted) line on which the neutralino and 
the stau masses are degenerate. 
The neutralino (Bino) is the lightest superparticle (LSP) within the MSSM particles in the region above 
the line. As the parameters approach to the line, the thermal abundance of the Bino LSP becomes 
consistent with the observed relic density of the cold dark matter, because the coannihilation works 
effectively. As a reference point which is displayed in Fig.~\ref{fig:mSUGRA_tb40_A0_0} by the blue 
point, a part of the mass spectrum of the SUSY particles is listed in Table~\ref{tab:mSUGRA} 
(C). We also show the SUSY contribution to the muon $(g-2)$ and the dark matter abundance by 
using the micrOMEGAs package~\cite{micromegas2.4}.

It is commented that the neutralino is lighter than the stau and hence the LSP in a wide region, e.g.
for small $m_0$, due to the large RGE contribution of the gauginos to the scalar masses.
This is contrary to the normal mSUGRA, and one of the notable features of the models with vector 
fields~\cite{Martin:2009bg}.

\begin{figure}[t]
\begin{center}
\includegraphics[width=8cm]{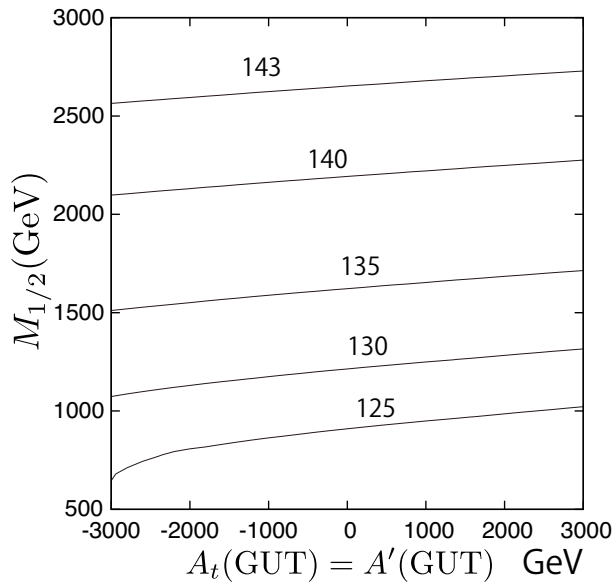}
\caption{Contours of the Higgs boson mass in $(A_t = A', M_{1/2})$ plane, where
the other parameters are $m_0 = M_{1/2}$ and $\tan\beta=40$. (The other $A$ parameters are set to be 0 at
the GUT scale.)
}
\label{fig:A_M12}
\end{center}
\end{figure}

The trilinear couplings were chosen $A_t = A' = 0$ at the GUT scale in Fig.~\ref{fig:mSUGRA_tb40_A0_0}.
As mentioned in Sec.~\ref{sec:RGE}, the result is insensitive to the choice because of the infrared 
fixed point behaviours. This feature is checked in Fig.~\ref{fig:A_M12}, where the contours of the 
Higgs boson are drawn. It is found that the Higgs boson mass depends dominantly on $M_{1/2}$ 
and is insensitive to the trilinear couplings. 

The maximal Higgs mass value might be enhanced without spoiling the muon $g-2$ prediction if the 
universal scalar masses are split at the GUT scale. However, it is found that the situation does not improve 
easily. Let us show the (incomplete) discussion of the non-universal mass spectrum. First, if the MSSM 
$\bf 10$ representation is heavier than that of $\bf \bar 5$, the Higgs mass tends to increase while the sleptons 
remain light. However, a heavy $\bf 10$, i.e. heavy stops, raises $\mu$, leading to a suppression of the 
muon $g-2$. Secondly, if the Higgs boson soft masses are taken to be non-universal and 
increased, the $\mu$ parameter decreases, 
which is favored by the muon $g-2$. However, since the stau tends to be tachyonic due to the Yukawa 
interaction during the RG evolution, the muon $g-2$ becomes suppressed as long as the soft scalar 
mass is universal among the generations. Anyway, this analysis is not complete, and the correlation 
between the Higgs boson mass and the muon $g-2$ may be relaxed by sophisticated models, which 
is an interesting topic for future. 

Summarizing the preceding study, the Higgs boson mass can be as large as about 130 GeV with the 
muon $g-2$ consistent with the experimental value at the 2$\sigma$ level for the mSUGRA boundary 
condition. The result is insensitive to the trilinear couplings in contrast to the usual mSUGRA. 
It is also mentioned that the Higgs boson mass can exceed 140 GeV if the muon $g-2$ were explained 
by other mechanism. As an example, a mass spectrum is listed at Table~\ref{tab:mSUGRA} (D).

\begin{table}[t]
\begin{center}
\begin{tabular}{|c|c|c|c|c|c|c|c|}
\hline
& $m_0$ & $M_{1/2}$ & $A_0$ & $\tan\beta$ & $M_{Q',U',E'}$
& $\Delta a_\mu$
& $m_h$ ($m_h^{\rm MSSM}$) 
\\ \hline
(C) 
& 60 & 1520 & 0 & 40 & 600
& $11.3 \times 10^{-10}$ 
& 129 (115) 
\\ \hline
(D) 
& 100 & 3500 & $-980$ & 40 & 600
& -- 
& 142 (117) 
\\ \hline
\end{tabular}

\vspace{1em}

\begin{tabular}{|c|c|c|c|c|c|c|c|c|c|}
\hline
& $m_{\tilde{g}}$ 
& $m_{\tilde{t}_1}$ & $m_{\tilde{T'}_{1-4}}$ 
& $\tilde{\mu}_{\rm L}$ & $\tilde{\mu}_{\rm R}$ & $\tilde{\tau}_1$ 
& $\chi^{\pm 1}_1$ 
& $\chi^0_1$ 
& $\Omega_{\chi^0_1} h^2$
\\ \hline
(C) 
& 1130 & 1469 & 1496--2014 
& 718 & 456 & 224 
& 320 & 202 
& 0.103
\\ \hline
(D) 
& 2759 & 3459 & 3258--4510
& 1692 & 1059 & 530
& 860 & 507
& 0.11
\\ \hline
\end{tabular}
\caption{A part of the mass spectrum and the value of deviation of the muon $g-2$,
for a mSUGRA model with vector-like fields, 
for a model point (C) in Fig.~\ref{fig:mSUGRA_tb40_A0_0}.
For comparison we have also shown the 
Higgs mass which would be obtained without the vector fields, as $m_h^{\rm MSSM}$.
All the masses are written in units of GeV.
The dark matter abundance $\Omega_{\chi^0_1} h^2$ is also shown.
For illustration, we also show a model point (D) with a Higgs mass above 140 GeV.
The neutralino contribution to the muon $g-2$ is comparable to that of the chargino at (C).
}
\label{tab:mSUGRA}
\end{center}
\end{table}

\section{Summary and Discussion}
\label{sec:summary}
We have investigated the muon $g-2$ and the Higgs boson mass
in a simple extension of the MSSM with extra vector-like matters.
As the mechanism of the SUSY breaking mediation,
both the GMSB models and mSUGRA have been studied.
\begin{itemize}
\item[(i)]
In GMSB models, the Higgs mass can be larger than 135 GeV (130 GeV) in the region where muon $g-2$ is consistent with the experimental value at the $2\sigma$ ($1\sigma$) level,
while maintaining the perturbative coupling unification. 
\item[(ii)]
In the case of mSUGRA models with universal soft masses, the Higgs mass can be as large as 
about 130 GeV when muon $g-2$ is within $2\sigma$ range.
Interestingly, the favored region overlaps with the coannihilation region of the neutralino dark matter scenario.
\end{itemize}
In both cases, the Higgs mass can be above 140 GeV if the muon $g-2$ constraint is not imposed.

Lastly, let us mention features of the collider phenomenology. In the parameter regions which are favored by the Higgs boson mass and the muon $g-2$, the SUSY particles are accessible by the LHC in near future. In addition, the scenario can be tested by direct searches of the fourth generations in the LHC.
A detailed study of LHC phenomenology of these models will be given elsewhere.

\section*{Acknowledgments}
This work was supported by Grand-in-Aid for Scientific research from
the Ministry of Education, Science, Sports, and Culture (MEXT), Japan,
No. 23740172 (M.E.), 
No. 21740164 (K.H.), No. 22244021 (K.H.) and No. 22-7585 (N.Y.).
S.I. is supported by JSPS Grant-in-Aid for JSPS Fellows.
This work was supported by World Premier International Research Center Initiative (WPI Initiative), MEXT, Japan.
K.H. and S.I. thank the YITP workshop ``Summer Institute 2011''.


\appendix

\section{Higgs mass correction from vector-like matters}
\label{app:mh}

The fermion mass matrix can be written as
\begin{eqnarray}
M_F = \left(
\begin{array}{cc}
M_{Q'} &Y' H_u^0 \\
0	& M_U'
\end{array}
\right),
\end{eqnarray}
and in the basis of $(\tilde{Q'} , \tilde{\bar{Q}'}^* , \tilde{\bar{U}'}, \tilde{{U}'}^*)^*\, M_S^2 \, (\tilde{Q'} , \tilde{\bar{Q}'}^* , \tilde{\bar{U}'}, \tilde{{U}'}^*)^T$, the scalar mass matrix is written as
\begin{eqnarray}
M_S^2 = \left(
\begin{array}{cccc}
|Y'H_u|^2 + m_{Q'}^2 + |M_{Q'}|^2 & (B_{Q'} M_{Q'})^* & (M_{U'}^* Y' H_u)^* & (A' Y' H_u-\mu^* Y' H_d^*)^* \\
(B_{Q'} M_{Q'}) & m_{\bar{Q}'}^2 + |M_{Q'}|^2 & 0 & (M_{Q'}^* Y' H_u)^* \\
M_{U'}^* Y' H_u & 0 & m_{\bar{U}'}^2 + |M_{U'}|^2 & (B_{U'} M_{U'})^* \\
(A' Y' H_u-\mu^* Y' H_d^*) & M_{Q'}^* Y' H_u & (B_{U'} M_{U'}) & |Y'H_u|^2+m_{{U}'}^2+ |M_{U'}|^2 
\end{array}
\right),
\end{eqnarray}
The one-loop corrections to the effective potential (in $\overline{\rm DR}$) can be written as
\begin{eqnarray}
\Delta' V = \frac{3}{32\pi^2}\left[ \sum_{i=1}^4 \tilde{m}_i^4 \left(\ln \frac{\tilde{m}_i^2}{Q^2}-\frac{3}{2}\right) -2 \sum_{i=1}^2 m_i^4 \left(\ln\frac{m_i^2}{Q^2} - \frac{3}{2}\right)\right],
\end{eqnarray}
where $\tilde{m}_i^2$ and $m_i^2$ are eigenvalues of $M_S^2$ and $M_F^\dag M_F$, respectively. The corrections to the soft mass of the up-type Higgs, $m_{H_u}^2$ is
\begin{eqnarray}
\Delta' m_{H_u}^2 = \left. \frac{1}{2 v_u } \frac{\partial \Delta' V}{\partial \left<H_u^0\right>} \right|_{\left<H_i^0\right>=v_i} .
\end{eqnarray}
\begin{eqnarray}
\frac{m_Z^2}{2} \simeq -|\mu|^2 -  (m_{H_u}^2 + \Delta m_{H_u}^2 ) + \mathcal{O}(1/\tan^{2}\beta),
\end{eqnarray}
where $\Delta m_{H_u}^2$ includes $\Delta' m_{H_u}^2$ and corrections from MSSM particles.

\section{Renormalization Group Equations}
\label{app:RGE}

In the numerical calculations, 
we used two-loop $\beta$ functions
for the renormalization group evolutions of the model parameters.
As emphasized in Ref.~\cite{Martin:2009bg}, the two loop effect is significant especially for the running of the gaugino masses.
In the $\overline{\rm DR}$ scheme, their explicit formulas are given by~\cite{twoloopRGE}:\footnote{We have used Susyno package~\cite{Fonseca:2011sy}.}
\begin{eqnarray}
16\pi^2\frac{dg_i}{dt} &=& b^{(1)}_i g_i^3 + \frac{g_i^3}{16\pi^2}
\left(\sum_{j=1}^3 b^{(2)}_{ij}g_j^2 - \sum_{x=t,b,\tau,4} c_{ix} Y_x^2\right)
\\
16\pi^2\frac{dM_i}{dt} &=& 2b^{(1)}_i g_i^2 M_i + \frac{2g_i^2}{16\pi^2}
\left(\sum_{j=1}^3 b^{(2)}_{ij}g_j^2(M_i+M_j)
 - \sum_{x=t,b,\tau,4} c_{ix} Y_x^2(M_i-A_x)\right)
\end{eqnarray}
where $t=\ln Q$, $Y_4=Y'$, $A_4=A'$, and
\begin{eqnarray}
&&
b_i^{(1)}=
\left(
\begin{array}{c}
\frac{33}{5}+3n_{10}
\\
1+3n_{10}
\\
-3+3n_{10}
\end{array}
\right)
\\
&&
b_{ij}^{(2)}=
\left(
\begin{array}{ccc}
\frac{199}{25}+\frac{23}{5}n_{10}
&
\frac{27}{5}+\frac{3}{5}n_{10}
&
\frac{88}{5}+\frac{48}{5}n_{10}
\\
\frac{9}{5}+\frac{1}{5}n_{10}
&
25+21n_{10}
&
24+16n_{10}
\\
\frac{11}{5}+\frac{6}{5}n_{10}
&
9+6n_{10}
&
14+34n_{10}
\end{array}
\right),
\;
c_{ix}=
\left(
\begin{array}{cccc}
\frac{26}{5} & \frac{14}{5} & \frac{18}{5} & \frac{26}{5}
\\
6 & 6 & 2 & 6
\\
4 & 4 & 0 & 4
\end{array}
\right)
\end{eqnarray}
In the above formulas,
$n_{10}=1$ corresponds to our setup, 
and $n_{10}=0$ and $Y'=0$ correspond to the MSSM case.

For completeness, we also show the contributions of the new fields and couplings 
to the $\beta$ functions of the other MSSM parameters:
\begin{eqnarray}
\frac{dX}{dt} &=& \left.\frac{dX}{dt}\right|_{\rm MSSM}
+\frac{1}{16\pi^2}\delta\beta^{(1)}(X)
+\frac{1}{(16\pi^2)^2}\delta\beta^{(2)}(X)\;.
\end{eqnarray}
The MSSM soft masses are denoted by
$m^2_{q_i}, m^2_{u_i}, m^2_{d_i}, m^2_{\ell_i}, m^2_{e_i}$ ($i=1\cdots 3$), $m_{H_u}^2$, and $m_{H_d}^2$, 
while $m^2_{Q'}, m^2_{\bar{Q}'}, m^2_{U'}, m^2_{\bar{U}'}, m^2_{E'}$, and $m^2_{\bar{E}'}$
are the soft masses for the vector fields.
\begin{eqnarray}
\delta\beta^{(1)}(Y_t) &=& 
3 Y'^2 Y_t 
\\
\delta\beta^{(2)}(Y_t) &=& 
Y_t \left(
\frac{13 g_1^4}{5}+9 g_2^4+16 g_3^4
+(\frac{4}{5} g_1^2 +16 g_3^2) Y'^2
-9 Y'^2 Y_t^2-9 Y'^4
\right)
\\
\delta\beta^{(1)}(Y_b) &=& 0
\\
\delta\beta^{(2)}(Y_b) &=& 
Y_b\left(
\frac{7 g_1^4}{5}+9 g_2^4+16 g_3^4-3 Y'^2 Y_t^2
\right)
\\
\delta\beta^{(1)}(Y_\tau) &=& 0
\\
\delta\beta^{(2)}(Y_\tau) &=& 
Y_\tau
\left(
\frac{27 g_1^4}{5}+9 g_2^4
\right)
\\
\delta\beta^{(1)}(\mu) &=& 
3 Y'^2 \mu
\\
\delta\beta^{(2)}(\mu) &=& 
\mu
\left(
\frac{9 g_1^4}{5}+9 g_2^4
+(\frac{4}{5} g_1^2 +16 g_3^2) Y'^2
-9 Y'^4
\right)
\\
\delta\beta^{(1)}(A_t) &=& 
6 A' Y'^2
%
\label{eq:betaAtop}
\\
\delta\beta^{(2)}(A_t) &=& 
-\frac{52}{5} g_1^4 M_1
-36 g_2^4 M_2
-64 g_3^4 M_3
+
\frac{8}{5} g_1^2 Y'^2 (A' - M_1)
+32 g_3^2 Y'^2 (A'  - M_3)
\nn && 
-18 Y'^2 Y_t^2 (A'+A_t)
-36 Y'^4 A' 
%
\\
\delta\beta^{(1)}(A_b) &=& 0
\\
\delta\beta^{(2)}(A_b) &=& 
-\frac{28}{5} g_1^4 M_1-36 g_2^4 M_2-64 g_3^4 M_3
-6 Y'^2 Y_t^2 (A'  + A_t)
%
\\
\delta\beta^{(1)}(A_\tau) &=& 0
\\
\delta\beta^{(2)}(A_\tau) &=& 
-\frac{108}{5} g_1^4 M_1-36 g_2^4 M_2
%
\\
\delta\beta^{(1)}(B) &=& 
6 A' Y'^2
\\
\delta\beta^{(2)}(B) &=& 
-\frac{36}{5} g_1^4 M_1
-36 g_2^4 M_2
+ \frac{8}{5} g_1^2 Y'^2 (A' - M_1)
+32 g_3^2 Y'^2 (A' - M_3)
-36 A' Y'^4
\nn 
%
\\
\delta\beta^{(1)}(m_{q_{1,2,3}}^2) &=& 
\frac{1}{5} g_1^2 S'
\\
\delta\beta^{(2)}(m_{q_{1,2}}^2) &=& 
\frac{2}{75}g_1^4
\left(m_{Q'}^2
-12 m_{U'}^2
+20 m_{\bar{U}'}^2
+21 m_{E'}^2
-15 m_{\bar{E}'}^2
+45 M_ 1^2\right)
\nn &&
+g_1^2 S'_1
+
3 g_2^4 S'_2
+
\frac{16}{3}g_3^4 S'_3
\nn
\delta\beta^{(2)}(m_{q_3}^2) &=& \delta\beta^{(2)}(m_{q_{1,2}}^2)
-6  Y'^2 Y_t^2 (X_t + X' + 2 A_t A')
%
\\
\delta\beta^{(1)}(m_{u_{1,2,3}}^2) &=& 
-\frac{4}{5} g_1^2 S'
\\
\delta\beta^{(2)}(m_{u_{1,2}}^2) &=& 
\frac{4}{75}g_1^4
\left(
3 m_ {Q'}^2
+5 m_ {\bar{Q}'}^2
+64 m_ {U'}^2
-12 m_ {E'}^2
+ 60 m_ {\bar{E}'}^2
+360 M_ 1^2\right)
\nn&&
-4 g_1^2 S'_1
+
\frac{16}{3} g_3^4 S'_3
\nn
\delta\beta^{(2)}(m_{u_3}^2) &=& \delta\beta^{(2)}(m_{u_{1,2}}^2)
-12 Y'^2 Y_t^2 (X_t + X' + 2 A_t A')
\\
\delta\beta^{(1)}(m_{d_{1,2,3}}^2) &=& 
\frac{2}{5} g_1^2 S'
\\
\delta\beta^{(2)}(m_{d_{1,2,3}}^2) &=& 
\frac{2}{75}g_1^4 
\left(
3 m_ {Q'}^2
+m_ {\bar{Q}'}^2
-16 m_ {U'}^2
+48 m_ {\bar{U}'}^2
+48 m_ {E'}^2
-24 m_ {\bar{E}'}^2
+180 M_ 1^2
\right)
\nn &&
+2 g_1^2 S'_1
+
\frac{16}{3} g_3^4 S'_3
%
\\
\delta\beta^{(1)}(m_{\ell_{1,2,3}}^2) &=& 
-\frac{3}{5} g_1^2 S'
\\
\delta\beta^{(2)}(m_{\ell_{1,2,3}}^2) &=& 
\frac{2}{25}
g_1^4
 \left(
 m_ {Q'}^2
 +2 m_ {\bar{Q}'}^2
 +28 m_ {U'}^2
  -4 m_ {\bar{U}'}^2
 -9 m_ {E'}^2
+ 27 m_ {\bar{E}'}^2
 +135 M_ 1^2\right)
\nn &&
-3 g_1^2 S'_1 
+ 3 g_2^4 S'_2
%
\\
\delta\beta^{(1)}(m_{e_{1,2,3}}^2) &=& 
\frac{6}{5} g_1^2 S'
\\
\delta\beta^{(2)}(m_{e_{1,2,3}}^2) &=& 
\frac{2}{25}g_1^4
\left(
7 m_ {Q'}^2
+5 m_ {\bar{Q}'}^2
+16 m_ {U'}^2
+80 m_ {\bar{U}'}^2
+ 72 m_ {e'}^2
+540 M_ 1^2\right)
+6 g_1^2 S'_1
%
\\
\delta\beta^{(1)}(m_{H_u}^2) &=& 
\frac{3}{5} g_1^2 S'
+6 Y'^2 X'
\label{eq:betaHu}
\\
\delta\beta^{(2)}(m_{H_u}^2) &=& 
\frac{2}{25}g_1^4
 \left(
2 m_ {Q'}^2
 +m_ {\bar{Q}'}^2
 -4 m_ {U'}^2
  +28 m_ {\bar{U}'}^2
 +27 m_ {E'}^2
-9 m_ {\bar{E}'}^2
 +135 M_ 1^2
 \right)
\nn &&
+
3 g_1^2 S'_1 
+
\frac{8}{5}g_1^2 Y'^2 (X' + 2M_1^2 - 2 M_1 A')
\nn &&
+ 3 g_2^4 S'_2
+ 32 g_3^2 Y'^2 (X' + 2M_3^2 - 2 M_3 A')
-36 Y'^4 (X' +  A'^2)
\\
\delta\beta^{(1)}(m_{H_d}^2) &=& 
-\frac{3}{5} g_1^2 S'
\\
\delta\beta^{(2)}(m_{H_d}^2) &=& 
\frac{2}{25}g_1^4
 \left(
 m_ {Q'}^2
 +2 m_ {\bar{Q}'}^2
 +28 m_ {U'}^2
  -4 m_ {\bar{U}'}^2
 -9 m_ {E'}^2
+ 27 m_ {\bar{E}'}^2
 +135 M_ 1^2\right)
\nn
&&
-3 g_1^2 S'_1
+3 g_2^4 S'_2
\end{eqnarray}
where 
\begin{eqnarray}
S' &=& 
m_{Q'}^2 - m_{\bar{Q'}}^2
-2m_{U'}^2 +2 m_{\bar{U'}}^2
+m_{E'}^2 - m_{\bar{E'}}^2
\\
S'_1 &=&
\frac{3}{5}g_2^2\left(
m_{Q'}^2 - m_{\bar{Q'}}^2
\right)
+
\frac{16}{15}g_3^2\left(
m_{Q'}^2 - m_{\bar{Q'}}^2 - 2 m_{U'}^2 + 2 m_{\bar{U}'}^2
\right)
\nn &&
+
\frac{2}{5}Y'^2\left(
-m_{Q'}^2+4m_{U'}^2-3m_{H_u}^2
\right)
\\
S'_2 &=& 3 m_{Q'}^2 + 3 m_{\bar{Q'}}^2 + 18 M_2^2
\\
S'_3 &=& 2 m_{Q'}^2 + 2 m_{\bar{Q'}}^2 + m_{U'}^2 + m_{\bar{U}'}^2 + 18 M_3^2
\\
X_t &=& m_{H_u}^2 + m_{q_3}^2 + m_{u_3}^2 + A_t^2
\\
X' &=& m_{H_u}^2 + m_{Q'}^2 + m_{U'}^2 + A'^2
\end{eqnarray}
In addition, the $\beta$-functions for the new parameters are given by
\begin{eqnarray}
\frac{dX'}{dt} &=& 
\frac{1}{16\pi^2}\beta^{(1)}(X')
+\frac{1}{(16\pi^2)^2}\beta^{(2)}(X')\;.
\end{eqnarray}
where
\begin{eqnarray}
\beta^{(1)}(Y') &=& 
Y'
\left(
-\frac{13 g_1^2}{15}-3 g_2^2-\frac{16 g_3^2}{3}+3 Y_t^2+6 Y'^2
\right)
%
\label{eq:betaY}
\\
\beta^{(2)}(Y') &=& 
Y'
\left(
\frac{3913 g_1^4}{450}+g_2^2 g_1^2+\frac{136}{45} g_3^2 g_1^2
+\frac{33 g_2^4}{2}+\frac{128 g_3^4}{9}+8 g_2^2 g_3^2
\right. \nn && \left.
+(\frac{4}{5} g_1^2+16 g_3^2) Y_t^2
+(\frac{6}{5} g_1^2 +6 g_2^2 +16 g_3^2) Y'^2
\right. \nn && \left.
-9 Y_t^4-9 Y'^2 Y_t^2-22 Y'^4-3 Y_b^2 Y_t^2
\right)
%
\\
\beta^{(1)}(A') &=& 
\frac{26}{15} g_1^2 M_1+6 g_2^2 M_2+\frac{32}{3} g_3^2 M_3
+ 6 Y_t^2 A_t 
+12 Y'^2 A' 
\label{eq:betaA}
%
\\
\beta^{(2)}(A') &=& 
-\frac{7826}{225} g_1^4 M_1
-66 g_2^4 M_2
-\frac{512}{9} g_3^4 M_3
\nn &&
-2 g_2^2 g_1^2 (M_1 + M_2)
-16 g_2^2 g_3^2 (M_2 + M_3)
-\frac{272}{45} g_1^2 g_3^2 (M_1 + M_3)
\nn &&
+\frac{8}{5} g_1^2 Y_t^2 (A_t - M_1)
+\frac{12}{5} g_1^2 Y'^2 (A'-M_1)
\nn &&
+12 g_2^2 Y'^2 (A' - M_2)
+32 g_3^2 Y_t^2 (A_t - M_3)
+32 g_3^2 Y'^2 (A' - M_3)
\nn &&
-6 Y_b^2 Y_t^2 (A_b + A_t)
-18 Y'^2 Y_t^2 (A' + A_t)
-36Y_t^4 A_t 
-88Y'^4 A' 
\\
\beta^{(1)}(B_{Q'}) &=& 
\frac{2}{15} g_1^2 M_1+6 g_2^2 M_2+\frac{32}{3} g_3^2 M_3 + 2 Y'^2 A' 
\\
\beta^{(2)}(B_{Q'}) &=& 
-\frac{578}{225} g_1^4 M_1-66 g_2^4 M_2-\frac{512}{9} g_3^4 M_3
\nn &&
-\frac{2}{5} g_1^2 g_2^2 (M_1 + M_2)
-32 g_2^2 g_3^2 (M_2 + M_3)
-\frac{32}{45} g_1^2 g_3^2 (M_1 + M_3)
\nn &&
+ \frac{8}{5} g_1^2 Y'^2 (A' - M_1)
-6 Y'^2 Y_t^2 (A' + A_t) -20 A' Y'^4
%
\\
\beta^{(1)}(B_{U'}) &=& 
\frac{32}{15} g_1^2 M_1+\frac{32}{3} g_3^2 M_3 + 4Y'^2 A' 
\\
\beta^{(2)}(B_{U'}) &=& 
-\frac{9728}{225} g_1^4 M_1
-\frac{512}{9} g_3^4 M_3
-\frac{512}{45} g_1^2 g_3^2 (M_1 + M_3)
\nn &&
-\frac{4}{5} g_1^2 Y'^2 (A' - M_1)
+12 g_2^2 Y'^2 (A' - M_2)
\nn &&
-12 Y'^2 Y_t^2 (A' + A_t)
-32 A' Y'^4 
%
\\
\beta^{(1)}(B_{E'}) &=& 
\frac{24}{5} g_1^2 M_1
\\
\beta^{(2)}(B_{E'}) &=& 
-\frac{2592}{25} g_1^4 M_1
%
\\
\beta^{(1)}(m_{Q'}^2) &=& 
\frac{1}{5}g_1^2 S
-\frac{2}{15} g_1^2 M_1^2
-6 g_2^2 M_2^2
-\frac{32}{3} g_3^2 M_3^2
+2 Y'^2 X'
\\
\beta^{(2)}(m_{Q'}^2) &=& 
\frac{1}{75}g_1^4
\left(
\sum(2 m_ {q_i}^2
-24 m_ {u_i}^2
+6 m_ {d_i}^2
-6 m_ {\ell_i}^2
+42 m_ {e_i}^2)
+12 m_ {H_u}^2
-6 m_ {H_d}^2
\right. \nn && \left.
+2 m_ {Q'}^2
-24 m_ {U'}^2
+40 m_ {\bar{U}'}^2
+42 m_ {E'}^2
-30 m_ {\bar{E}'}^2
+289 M_ 1^2
\right)
\nn &&
+ g_1^2 S_1
+ 3 g_2^4 S_2 
+ \frac{16}{3}g_3^4 S_3 
+\frac{2}{5}g_1^2 g_2^2 (M_1^2 + M_2^2 + M_1  M_2)
\nn &&
+\frac{32}{45}g_1^2 g_3^2 (M_1^2 + M_3^2 + M_1  M_3)
+ 32 g_2^2 g_3^2 (M_2^2 + M_3^2 + M_2 M_3)
\nn &&
+ \frac{8}{5}g_1^2 Y'^2 (X' + 2 M_1^2 - 2 M_1 A')
\nn &&
-20 Y'^4 (X' + A'^2)
-6Y_t^2 Y'^2 (X' + X_t + 2 A'A_t)
%
\\
\beta^{(1)}(m_{\bar{Q}'}^2) &=& 
-\frac{2}{15} g_1^2 M_1^2-6 g_2^2 M_2^2-\frac{32}{3} g_3^2 M_3^2-\frac{g_1^2 S}{5}
\\
\beta^{(2)}(m_{\bar{Q}'}^2) &=& 
\frac{1}{75}g_1^4\left(
\sum(40 m_ {u_i}^2
-2 m_ {d_i}^2
+12 m_ {\ell_i}^2
-30 m_ {e_i}^2)
-6 m_ {H_u}^2
+ 12 m_ {H_d}^2
\right. \nn && \left.
+2 m_ {\bar{Q}'}^2
+40 m_ {U'}^2
-24 m_ {\bar{U}'}^2
-30 m_ {E'}^2
+42 m_ {\bar{E}'}^2
+289 M_ 1^2
\right)
\nn &&
-g_1^2 S_1
+3g_2^2 S_2
+\frac{16}{3}g_3^2 S_3
+\frac{2}{5}g_1^2 g_2^2 (M_1^2 + M_2^2 + M_1  M_2)
\nn &&
+\frac{32}{45}g_1^2 g_3^2 (M_1^2 + M_3^2 + M_1  M_3)
+ 32 g_2^2 g_3^2 (M_2^2 + M_3^2 + M_2 M_3)
\\
\beta^{(1)}(m_{U'}^2) &=& 
-\frac{32}{15} g_1^2 M_1^2-\frac{32}{3} g_3^2 M_3^2-\frac{4 g_1^2 S}{5}
+ 4 Y'^2 X'
\\
\beta^{(2)}(m_{U'}^2) &=& 
\frac{4}{75}g_1^4
 \left(
 \sum(
 3 m_ {q_i}^2
 +64 m_ {u_i}^2
 +4 m_ {d_i}^2
  +21 m_ {\ell_i}^2
 -12 m_ {e_i}^2)
 +3 m_ {H_u}^2
+ 21 m_ {H_d}^2
\right. \nn && \left.
 +3 m_ {Q'}^2
 +5 m_ {\bar{Q}'}^2
  +64 m_ {U'}^2
 -12 m_ {E'}^2
 +60 m_ {\bar{E}'}^2
 +1216 M_ 1^2\right)
\nn &&
-4g_1^2 S_1
+ \frac{16}{3}g_3^2 S_3
+\frac{512}{45}g_1^2 g_3^2 (M_1^2 + M_3^2 + M_1  M_3)
\nn &&
-\frac{4}{5}g_1^2 Y'^2 (X' + 2 M_1^2 - 2 M_1 A')
+ 12 g_2^2 Y'^2 (X' + 2M_2^2 - 2M_2 A')
\nn &&
-32 Y'^2 (X' + A'^2)
-12 Y_t^2 Y'^2 (X' + X_t + 2 A' A_t)
\\
\beta^{(1)}(m_{\bar{U}'}^2) &=& 
-\frac{32}{15} g_1^2 M_1^2-\frac{32}{3} g_3^2 M_3^2+\frac{4 g_1^2 S}{5}
\\
\beta^{(2)}(m_{\bar{U}'}^2) &=& 
\frac{4}{75}g_1^4
 \left(
 \sum(
5 m_ {q_i}^2
 +12 m_ {d_i}^2
 +3 m_ {\ell_i}^2
 +60 m_ {e_i}^2)
 +21 m_ {H_u}^2
 + 3 m_ {H_d}^2
 \right. \nn && \left.
 +5 m_ {Q'}^2
 +3 m_ {\bar{Q}'}^2
  +64 m_ {\bar{U}'}^2
 +60 m_ {E'}^2
 -12 m_ {\bar{E}'}^2
 +1216 M_ 1^2\right)
 \nn &&
+ 4 g_1^2 S_1
+ \frac{16}{3} g_3^2 S_3
+ \frac{512}{45}g_1^2 g_3^2 (M_1^2 + M_3^2 + M_1  M_3)
\\
\beta^{(1)}(m_{E'}^2) &=& 
\frac{6 g_1^2 S}{5}-\frac{24}{5} g_1^2 M_1^2
\\
\beta^{(2)}(m_{E'}^2) &=& 
\frac{2}{25}g_1^4
 \left(
 \sum(
  7 m_ {q_i}^2
 +16 m_ {u_i}^2
 +16 m_ {d_i}^2
 +9 m_ {\ell_i}^2
 +72 m_ {e_i}^2)
 +27 m_ {H_u}^2
 + 9 m_ {H_d}^2
\right. \nn && \left.
 +7 m_ {Q'}^2
 +5 m_ {\bar{Q}'}^2
 +16 m_ {U'}^2
 +80 m_ {\bar{U}'}^2
 +72 m_ {E'}^2
 +1944 M_ 1^2\right)
+ 6 g_1^2 S_1 
\\
\beta^{(1)}(m_{\bar{E}'}^2) &=& 
-\frac{24}{5} g_1^2 M_1^2-\frac{6 g_1^2 S}{5}
\\
\beta^{(2)}(m_{\bar{E}'}^2) &=& 
\frac{2}{25}g_1^4
 \left(
\sum(5 m_ {q_i}^2
 +80 m_ {u_i}^2
 +8 m_ {d_i}^2
 +27 m_ {\ell_i}^2)
 +9 m_ {H_u}^2
+ 27 m_ {H_d}^2
\right. \nn && \left.
 +5 m_ {Q'}^2
 +7 m_ {\bar{Q}'}^2
 +80 m_ {U'}^2
  +16 m_ {\bar{U}'}^2
 +72 m_ {\bar{E}'}^2
 +1944 M_ 1^2\right)-6g_1^2 S_1
\end{eqnarray}
where
\begin{eqnarray}
S 
&=& 
m_{H_u}^2 - m_{H_d}^2 + \sum(m_{q_i}^2 -m_{\ell_i}^2 -2 m_{u_i}^2 + m_{d_i}^2 + m_{e_i}^2) + S'
\\
S_1 &=&
\frac{3}{5}g_2^2\left(
\sum (m_{q_i}^2-m_{\ell_i}^2 )
+m_{H_u}^2
-m_{H_d}^2 
\right)
+\frac{16}{15} g_3^2 \sum
\left(
m_{q_i}^2
- 2 m_{u_i}^2
+ m_{d_i}^2
\right)
\nn &&
+
\frac{2}{5}Y_t^2 
\left(
-3 m_{H_u}^2
- m_{q_3}^2 
+ 4 m_{u_3}^2
\right)
+
\frac{2}{5}Y_b^2
\left(
3 m_{H_d}^2
- m_{q_3}^2
- 2 m_{d_3}^2
\right) 
\nn &&
+
\frac{2}{5} Y_{\tau}^2
\left(
m_{H_d}^2
+ m_{L_3}^2 
- 2 m_{e_3}^2
\right)
+
S'_1
\\
S_2 &=& 
\sum(3  m_{q_i}^2 +  m_{\ell_i}^2) + m_{H_u}^2 + m_{H_d}^2
+ 11 M_2^2 + S'_2
\\
S_3 &=& 
\sum(2  m_{q_i}^2 + m_{u_i}^2 +  m_{d_i}^2)
-8 M_3^2 + S'_3
\end{eqnarray}




\end{document}